\documentclass[10pt,prb,aps,twocolumn,showpacs,
superscriptaddress]{revtex4}
\usepackage{graphicx}
\begin{document}

\title{Peierls transition in the presence of finite-frequency phonons in the
one-dimensional extended Peierls-Hubbard model at half-filling}

\author{Pinaki Sengupta} 
\affiliation{Department of Physics, University of California,
Davis, California 95616}

\author{Anders W. Sandvik} 
\affiliation{Department of Physics, {\AA}bo Akademi University, 
Porthansgatan 3, FIN-20500, Turku, Finland}

\author{David K. Campbell}
\affiliation{Departments of Physics and of Electrical and Computer 
Engineering, Boston University, 44 Cummington Street, Boston, 
Massachusetts 02215}

\date{\today}

\begin{abstract}
We report quantum Monte Carlo (stochastic series expansion) results for the 
transition from a Mott insulator to a dimerized Peierls insulating state in 
a half-filled, 1D extended Hubbard model coupled to optical bond phonons. 
Using electron-electron (e-e) interaction parameters corresponding 
approximately to polyacetylene, we show that the Mott-Peierls transition 
occurs at a finite value of the electron-phonon (e-ph) coupling. We discuss 
several different criteria for detecting the transition and show that they 
give consistent results. We calculate the critical e-ph coupling as a function
of the bare phonon frequency and also investigate the sensitivity of the 
critical coupling to the strength of the e-e interaction. In the limit of 
strong e-e couplings, we map the model to a spin-Peierls chain and compare 
the phase boundary with previous results for the spin-Peierls transition. 
We point out effects of a nonlinear spin-phonon coupling neglected in the 
mapping to the spin-Peierls model.
\end{abstract}


\maketitle

\section{Introduction}

Almost half a century ago, Peierls demonstrated that a one-dimensional (1D) 
metal coupled to an elastic lattice could exhibit an instability towards a 
lattice distortion of wave vector $q=2k_F$.\cite{peierls} This leads to a 
gap in the electronic spectrum at the Fermi energy and, for the case of a
half-filled band, the ground state is dimerized. Experimentally, the Peierls 
instability can be observed in a wide range of quasi-1D materials, e.g., 
conjugated polymers,\cite{pa} organic charge transfer salts,\cite{cts}
MX salts,\cite{mx} and $\mbox{CuGeO}_3$.\cite{cugeo3} To explain
quantitatively the properties of these materials, several different models 
extending beyond the original non-interacting Peierls model with a classical 
lattice have been proposed.\cite{review} These include the 
Su-Schrieffer-Heeger (SSH) model,\cite{ssh,nakahara,su,fradkin-ssh,zheng} 
the Holstein model,\cite{holstein,fradkin-hol,jeckelmann-hol} various 
Peierls-Hubbard \cite{phmodels,takahashi,fehske} and extended Peierls-Hubbard 
\cite{hirsch,zimanyi,takahashi,jeckelmann-mi,barford} models, 
as well as spin-Peierls models.
\cite{pincus,pytte,cross,sf,uhrig,gros,wellein,anders-sp,robert-sp,weisse,raas}

The Peierls instability is well understood in the static lattice limit 
(adiabatic phonons) in the absence of electron-electron (e-e) interactions.
The ground state is a Peierls state for arbitrarily small electron-phonon 
(e-ph) coupling. However, the quantum lattice and e-e interaction effects are 
still not completely understood. For many quasi-1D materials the zero-point 
fluctuations of the phonon field are comparable to the amplitude of the 
Peierls distortion,\cite{su} and this has spurred a large number of studies 
of quantum fluctuations in SSH,\cite{nakahara,fradkin-ssh,zheng} and Holstein,
\cite{fradkin-hol,jeckelmann-hol} models. Recent numerical studies have shown 
that quantum fluctuations destroy the Peierls instability for small 
e-ph  coupling and/or large phonon frequency in both the spinless
\cite{sf} and spin-$\frac{1}{2}$ Holstein models \cite{jeckelmann-hol} at 
half-filling. For the SSH model, Fradkin and Hirsch\cite{fradkin-ssh}
carried out an extensive study of spin-${1 \over 2}$ ($n=2$) and spinless 
($n=1$) fermions. In the anti-adiabatic limit (vanishing ionic mass), 
they mapped the system to an $n$-component Gross-Neveu model, which is known 
to have long-ranged dimerization for arbitrary coupling for $n\geq 2$ but 
not for $n=1$. For non-zero ionic mass, a renormalization group 
analysis showed that the low-energy behavior of the $n=2$ model is still 
governed by the zero mass limit of the theory. This implies that the 
spin-${1 \over 2}$ SSH model (but not the spinless model) has a dimerized 
ground state for arbitrarily weak e-ph coupling. Early numerical calculations
are also consistent with this scenario.\cite{fradkin-ssh} It should be noted,
however, that a dimerized state was also predicted for any non-zero e-ph 
coupling in the spin-${1 \over 2}$ Holstein model,\cite{fradkin-hol} for 
which more recent large-scale calculations have instead indicated a non-zero 
critical coupling.\cite{jeckelmann-hol}

Independent-electron models, such as SSH and Holstein, are important
from a theoretical standpoint but are not sufficient to account quantitatively 
for the experimentally observed properties of real materials. For 
that e-e interactions have to be included in addition to the e-ph couplings.
\cite{pa} The interplay among the different interactions gives rise to a 
rich variety of broken-symmetry ground states as well as low-energy 
electron-lattice excitations like solitons, polarons, bipolarons, etc.
\cite{pa} At half-filling, on-site (Hubbard) interactions open a charge
gap,\cite{liebwu} and in the absence of e-ph couplings the system is then a 
Mott insulator with algebraically decaying ($1/r$ as a function of distance 
$r$) spin-spin correlations. Hence, the Peierls transition in this case is  
accompanied only by the opening of a spin gap, the charge gap already 
generated by the e-e interactions. Longer-range e-e interactions can destroy 
the Mott state, however, and hence must affect also the Mott-Peierls 
transition. Without phonons, even in the simplest half-filled extended Hubbard
model with only on-site ($U$) and nearest-neighbor ($V$) interactions,
\cite{solyom,voit} some features of the phase diagram are still controversial.
\cite{nakamura,pinaki,jeckelman-ehm,jcomment} Adding e-ph interactions 
further increases the complexity of the problem, and the determination of 
the phase diagram remains a very challenging problem.

Zimanyi et al.\cite{zimanyi} have investigated models with both e-e and e-ph 
interactions using ``g-ology'' and RG techniques. They showed that the 
ground state has a spin gap if the combined backscattering 
amplitude $g_1^T=g_1(\omega_0)+\tilde{g}_1(\omega_0)$ is negative, where 
$g_1(\omega_0)$ is the contribution from e-e interactions and $\tilde{g}_1
(\omega_0) < 0$ comes from the e-ph interactions. Thus, in the extended 
Hubbard model, if the bare coupling $g_1=U-2V$ is positive, then $g_1^T 
\geq 0$ for small values of the e-ph coupling and there is no spin gap. The 
transition to a Peierls state occurs only when the e-ph coupling exceeds a 
critical value and $g_1^T$ becomes negative. It should be noted, however, 
that the conventional scenario for the behavior close to the line $U=2V$ 
has recently been challenged \cite{nakamura,pinaki,jcomment} and a 
Peierls-like bond-ordered state most likely appears close to $V=U/2$ 
{\it even in the absence of e-ph couplings}. For the pure SSH model ($U=V=0$),
$g_1$ is zero and $g_1^T$ is negative for any non-zero e-ph coupling. This 
implies a Peierls ground state for arbitrarily small e-ph coupling, in 
agreement with the earlier results of Fradkin and Hirsch.\cite{fradkin-ssh} 

In the limit of strong on-site e-e interactions, which inhibit doubly 
occupied sites, a half-filled system can be mapped to a spin-phonon model, 
which also can undergo a dimerization (spin-Peierls) transition.
\cite{pincus,cross} Extensive studies of spin-Peierls models in recent years 
\cite{uhrig,gros,wellein,anders-sp,robert-sp,weisse,raas} have largely been 
spurred by the discovery of a spin-Peierls transition at unusually high 
temperature (14~K) in $\mbox{CuGeO}_3$.\cite{cugeo3} Several different 
calculations, for different types of spin-phonon couplings, have shown that 
the transition occurs only above a finite spin-phonon coupling in the 
presence of 
finite-frequency phonons.\cite{uhrig,anders-sp,robert-sp,weisse,raas} This
is in contrast to the adiabatic limit, where dimerization occurs for 
infinitesimal coupling.\cite{cross} 

Numerical studies of models with both e-ph and e-e interactions, in which 
the charge degrees of freedom are retained (Peierls-Hubbard and extended 
Peierls Hubbard models), have addressed the effect of interactions on the 
dimerization amplitude \cite{hirsch,takahashi} and the excited states.
\cite{barford} Detailed studies of the phase diagrams have in the past been 
limited by the small lattice sizes accessible when both e-e and e-ph 
interactions are included.\cite{fehske} The situation is rapidly improving, 
however, as modern quantum Monte Carlo \cite{sseloop,anders-sp,pinaki} and 
density matrix renormalization group \cite{robert-sp,barford} methods can now
access models with both e-e and e-ph interactions on chains with several 
hundred sites.

Here we consider a 1D extended Hubbard model with on-site ($U$) and 
nearest-neighbor ($V$) e-e interactions and couple it to dispersionless 
optical bond phonons via modulation of the electron kinetic energy. We study 
the transition from a Mott insulating state, with dominant spin-spin 
correlations, to a Peierls (dimerized) spin-gapped state. Since 
the parameter space of this model is rather large, with a bare phonon 
frequency ($\omega_0$) and an e-ph coupling ($\alpha$) in addition to the 
e-e couplings, we have limited our study to a physically reasonable ratio 
$U/V=4$ of the e-e parameters. 

In Sec.~II we define the model and the various 
physical quantities that we have calculated using a quantum Monte Carlo method 
(stochastic series expansion \cite{sseloop}). In Sec.~III we discuss several 
signals that we have used to detect the Mott-Peierls transition. In Sec.~IV we
present the phase diagram in the ($\omega_0,\alpha$)-plane for a fixed value 
of the on-site interaction $U$ that has previously been used in models of 
polyacetylene. We also discuss the effects of varying the e-e interaction 
strength at fixed $\omega_0$. We map the model to a spin-Peierls model for 
large $U,V$ and compare the Mott-Peierls boundary with known spin-Peierls 
results. In Sec.~V we summarize our results and discuss some future prospects.

\section{Model and observables}

The Hamiltonian is given by
\begin{eqnarray}
H &=& -t\sum_{i,\sigma}(1+\alpha[a^{\dagger}_i + a_i])
(c^{\dagger}_{i+1,\sigma}c_{i,\sigma} + c^{\dagger}_{i,\sigma} c_{i+1,\sigma})
\nonumber \\
& & + \mu\sum_in_i + U\sum_i(n_{i,\uparrow}-{1\over 2})
(n_{i,\downarrow}-{1\over 2}) \nonumber \\
  & & + V\sum_i(n_i - 1)(n_{i+1} - 1) + \omega_0\sum_ia^{\dagger}_ia_i,
\label{eqn:H}
\end{eqnarray}
where $a^{\dagger}_i(a_i)$ creates(annihilates) a phonon on the bond
between sites $i$ and $i+1$, $c^{\dagger}_{i,\sigma}(c_{i,\sigma})$ is the 
spin-$\sigma$ electron creation(annihilation) operator, and $n_i = 
n_{i,\downarrow}+n_{i,\uparrow}$. For the half-filled band that we study here,
the chemical potential $\mu=0$. We set the single-electron hopping $t$
to unity. For the e-e interactions, we first take the values $U=2.5$, $V=U/4$,
which have previously been used as values corresponding approximately to 
what is expected in polyacetylene.\cite{jeckelmann-mi} We will also consider 
other values of $U$, keeping the ratio fixed at $U/V=4$. We study the system 
as a function of the bare phonon frequency $\omega_0$ and the e-ph interaction
$\alpha$. 

The dispersionless optical phonons we use are different from the bare SSH 
phonons, which have vanishing energy for momentum $q\to 0$. However, these 
acoustic phonons decouple from the electronic low-energy states involved
in the Peierls instability and therefore only the optical phonons close to 
$q=\pi$ need to be kept.\cite{fradkin-ssh,zimanyi} Hence, in this regard we 
expect the optical phonons in (\ref{eqn:H}) to be equivalent to fully quantum 
mechanical SSH phonons. In the non-interacting limit ($U,V \rightarrow 0$), 
the ground state should therefore be a dimerized Peierls insulator for any 
non-zero $\alpha$.\cite{fradkin-ssh}

To obtain numerically exact ground state results we have used the 
stochastic series expansion (SSE) quantum Monte Carlo method \cite{sseloop}
for periodic chains with up to $N=256$ sites. The SSE method is a 
finite-temperature technique based on importance sampling of the diagonal 
elements of the Taylor expansion of $e^{-\beta H}$, where $\beta$ is the 
inverse temperature; $\beta=t/T$. Ground state expectation values can be 
obtained using sufficiently large $\beta$, and there are then no 
approximations beyond the statistical errors. Typically, $\beta=2N$ or $4N$ 
was sufficient for the quantities presented here to have converged to their 
ground state values. Using the recently developed ``operator loop'' update,
\cite{sseloop} the electronic degrees of freedom are treated in the same 
manner as in the absence of phonons.\cite{pinaki} The phonons are also
treated in the occupation number basis directly with the SSE representation 
\cite{pinakithesis} (i.e., slightly different from the interaction picture
used for the phonons in in Refs.~\onlinecite{anders-sp} and 
\onlinecite{irsse}). At the (low) energy scales that we are interested in
here, the number of phonons per bond is small (typically $<10$) and there are 
no problems in using a truncated basis (the truncation can be arbitrarily 
large in the SSE).

The Mott and Peierls phases can be characterized using the static spin (S)
and bond (B) structure factors and susceptibilities, as will be further
discussed in Sec.~III.  The structure factors are defined by
\begin{eqnarray}
S_S(q) &=& \frac{1}{N} \sum_{k,l}e^{iq(k-l)}\langle
S_k^zS_l^z\rangle, \label{eq:ssdw} \\
S_B(q) &=& \frac{1}{N} \sum_{k,l}e^{iq(k-l)}
\langle K_kK_l\rangle, 
\label{eq:sbow}
\end{eqnarray}
\noindent where 
$K_j=\sum_{\sigma=\uparrow,\downarrow}(c^{\dagger}_{j+1,\sigma}c_{j,\sigma} 
+ {\rm h.c.})$ is the kinetic energy operator on the $i$th bond. The 
corresponding static susceptibilities are given by
\begin{eqnarray}
\chi_S(q) & = & \frac{1}{N} \sum_{k,l}e^{iq(k-l)}\int^\beta_0 d\tau \langle 
S^z_k(\tau) S^z_l(0)\rangle, \label{eq:xsdw} \\
\chi_B(q) & = & \frac{1}{N} \sum_{k,l}e^{iq(k-l)}\int^\beta_0 d\tau 
\langle K_k(\tau) K_l(0)\rangle.
\label{eq:xbow}
\end{eqnarray}
Direct evidence for the presence or absence of the spin and charge
gaps can also be obtained from spin and charge stiffnesses $\rho_c$ and
$\rho_s$, which are defined as the second derivative of the internal energy 
per site with respect to phase factors multiplying the kinetic energy; 
$\rho_{c,s} = \partial ^2E(\phi_{c,s})/\partial\phi_{c,s} ^2$.\cite{kohn}
The SSE estimators for all these observables can be found in 
Ref.~\onlinecite{pinaki}.

\section{Detecting the Mott-Peierls transition}

\begin{figure}
\includegraphics[width=8.3cm]{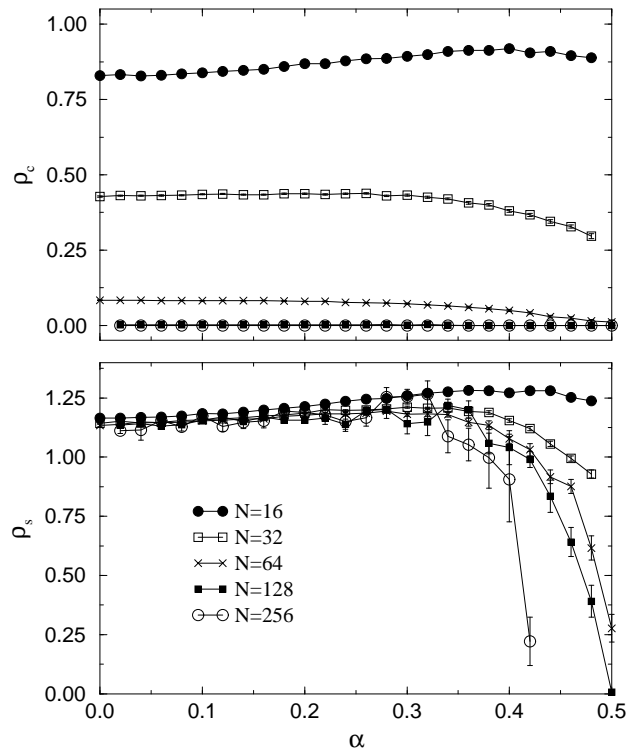}
\caption{Charge (upper panel) and spin (lower panel) stiffness constants
vs e-ph coupling for several different system sizes at $\omega_0=1$ and 
$U=2.5$.} 
\label{fig:rho}
\end{figure}
  
For our choice of $V=U/4$, the ground state in the limit of zero 
e-ph interaction ($\alpha=0$) is a Mott insulator with no spin gap (but 
finite charge gap) and is characterized by a $1/r$ decay of the staggered 
spin-spin correlations.\cite{voit,pinaki} The transition to a dimerized 
Peierls state is marked by the development of a staggered kinetic energy
modulation and will hence be signalled by divergent peaks at $q=\pi$ in 
the bond-order structure factor, Eq.~(\ref{eq:sbow}), and susceptibility, 
Eq.~(\ref{eq:xbow}). The dimerization is accompanied by the opening of a 
spin gap, the charge gap remaining finite. Hence  the $q=\pi$ peak in the 
spin structure factor and susceptibility, Eqs.~(\ref{eq:ssdw}) and 
(\ref{eq:xsdw}), which diverge in the Mott phase, become non-divergent
in the Peierls state. 

In the adiabatic limit, the system is dimerized for any $\alpha>0$. We here 
present several results showing that the Peierls transition occurs at a 
critical coupling $\alpha_c>0$ when $\omega_0=1$ and $U=2.5$. The phase 
diagrams discussed in Sec.~IV are based on the same signals for the transition
at other e-e couplings and $\omega_0$.

Since the charge gap is finite in both the Mott and Peierls states, the charge 
stiffness $\rho_c$ should vanish in the thermodynamic limit for all $\alpha$.
The upper panel of Fig.~\ref{fig:rho} shows $\rho_c$ as a function of $\alpha$
for several system sizes $N$. As $N$ grows, $\rho_c$ indeed rapidly converges
to zero for all $\alpha$, in agreement to our expectations. The Mott state has 
no spin gap (finite spin stiffness) whereas the Peierls state has a finite 
spin gap (zero spin stiffness). If the Peierls transition occurs at a 
critical coupling $\alpha_c>0$, it should be of the Kosterliz-Thouless type, 
\cite{voit} where the spin stiffness changes discontinuously from a finite 
value for $\alpha\le\alpha_c$ to zero for $\alpha>\alpha_c$. The spin 
stiffness graphed in the lower panel of Fig.~\ref{fig:rho} shows a jump
developing with increasing $N$, indicating a critical coupling
$\alpha_c \approx 0.3$.

\begin{figure}
\includegraphics[width=8.3cm]{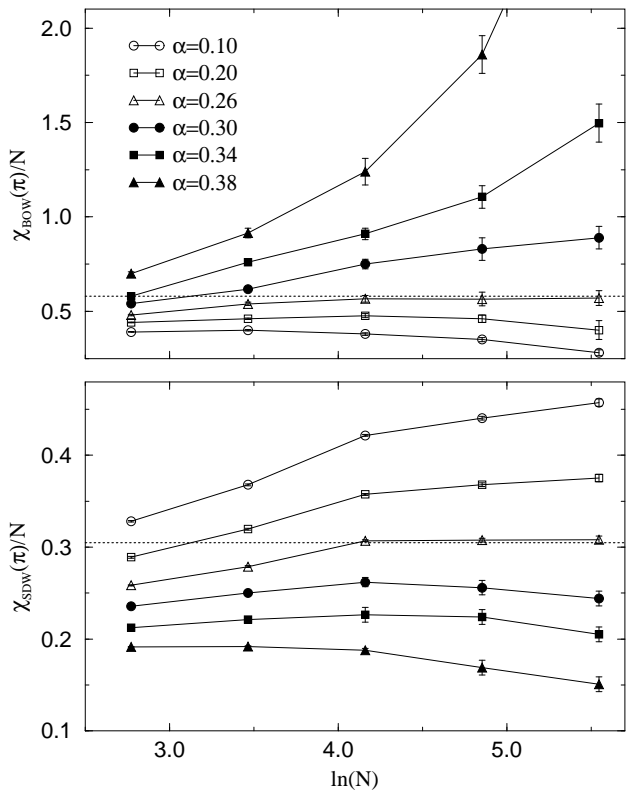}
\caption{Finite-size scaling of the staggered bond-order (upper panel)
and spin (lower panel) susceptibility for $\omega_0/t=1$ and several values 
of $\alpha$. The dashed lines show the $N$-independent behavior expected 
att the Mott-Peierls transition.} 
\label{fig:xbow}
\end{figure}

The spin stiffness data do not easily yield a more accurate estimate of the 
critical coupling. As discussed in the context of the spin-Peierls model, 
\cite{anders-sp} logarithmic corrections lead to large finite-size effects 
for $\alpha \approx \alpha_c$. A more accurate estimate can be 
obtained from the scaling behavior of the finite-size staggered bond and 
spin susceptibilities.\cite{anders-sp} It is known from bosonization studies
that in the Mott phase the equal-time staggered spin and bond 
correlations both decay with distance $r$ as $1/r$, up to multiplicative 
logarithmic corrections.\cite{voit} At the Mott-Peierls phase boundary, the 
log-corrections can be expected to disappear,\cite{eggert1} and this can be 
used as a criterion for the phase transition. In the dimerized Peierls phase 
the bond correlation function
approaches a constant at long distances, whereas the spin
correlations decay exponentially. It is convenient to study the associated 
static susceptibilities defined in Eqs.~(\ref{eq:xsdw}) and (\ref{eq:xbow}), 
which in a critical state scale with one power of $N$ higher than the 
structure factors Eqs.~(\ref{eq:ssdw}) and (\ref{eq:sbow}). In the 
Peierls phase $\chi_S(\pi)/N$ should converge to $0$ and $\chi_B(\pi)/N$ 
should diverge, whereas in the Mott phase $\chi_S(\pi)/N$ should diverge 
logarithmically and $\chi_B(\pi)/N$ should approach zero logarithmically (the 
log-corrections for spin and bond correlations are different\cite{voit}). 
Fig.~\ref{fig:xbow} shows both quantities versus $\mbox{ln}(N)$ for several 
values of $\alpha$. The expected behavior is indeed observed, and within 
statistical errors both $\chi_S(\pi)/N$ and $\chi_B(\pi)/N$ are independent 
of $N$ for the largest chains when $\alpha \approx 0.26$.

\begin{figure}
\includegraphics[width=8.3cm]{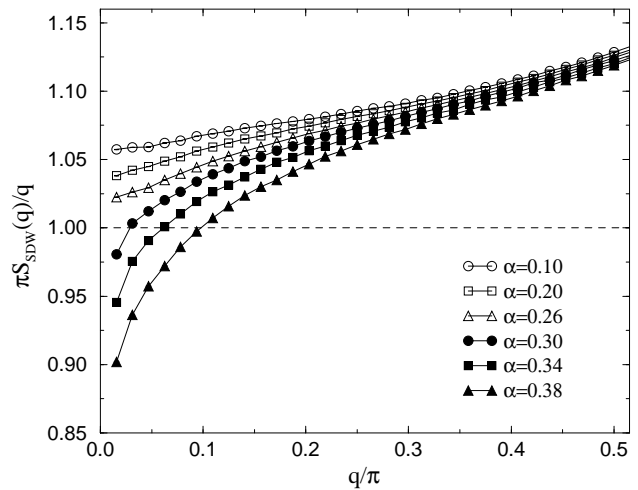}
\caption{$S_S(q)/q$ vs $q$ for several values of $\alpha$ 
around $\alpha_c$ for $N$=128. The curves for $\alpha \ge0.30$ dip below 
$1$ for small $q$, indicating the presence of a spin-gap.} 
\label{fig:sq}
\end{figure}

Additional confirmation of the critical coupling is obtained by studying the 
behavior of $S_S(q)/q$ as $q\rightarrow 0$. It has been shown 
\cite{voit,torsten} 
that if the ground state is gapless in the spin sector, 
then $S_S(q)/q\rightarrow 1/\pi$ as $q\rightarrow 0$, 
whereas if there is a spin gap, $S_S(q)/q \rightarrow 0$. Even a very small 
spin gap can be detected this way, since it is in practice sufficient to see 
that $\pi S_S(q)/q$ decays below $1$ for small $q$ to conclude that a 
spin gap must be present. In Fig.~\ref{fig:sq} we present results for 
different values of $\alpha$. The curves for $\alpha \le 0.26$ are above 
1 for all $q$, and the decay towards 1 is very slow. The asymptotic
approach to 1 can be expected to be logarithmic.\cite{eggert2} On the other 
hand, the $\alpha \ge 0.30$ curves drop below 1. From these results
we estimate $\alpha_c$=0.28$\pm$0.01, which is compatible with the 
$q=\pi$ quantities in Fig.~\ref{fig:xbow}. In general, we have found that 
$S_S(q)/q$, which indirectly signals the opening of a spin gap, is the 
easiest and most reliable way to detect a bond-ordered state (see also 
Ref.~\onlinecite{pinaki}). 

\section{phase diagrams}

\begin{figure}
\includegraphics[width=8.3cm]{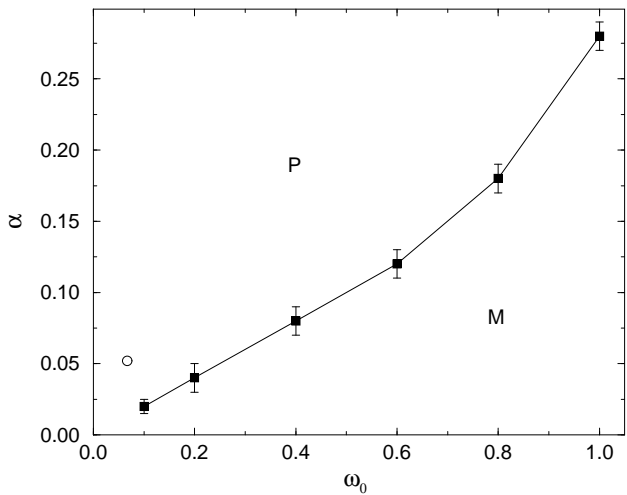}
\caption{Phase diagram for $U$=2.5, $V$=0.625. The squares with error bars 
show the critical e-ph coupling separating the Mott (M) and Peierls (P) 
insulating phases. The circle corresponds to phonon parameters previously
used for polyacetylene.\cite{fradkin-ssh}}
\label{fig:alphac}
\end{figure}

The critical coupling depends on the parameters of the Hamiltonian, in 
particular, the bare phonon frequency $\omega_0$. Using the above criteria 
for distinguishing the Mott and Peierls phases, we have calculated $\alpha_c$ 
as a function of $\omega_0$, keeping $U=2.5$, $V=0.625$. As shown in 
Fig.~\ref{fig:alphac}, $\alpha_c$ decreases linearly to zero for small 
$\omega_0$. This phase diagram is hence consistent with the known 
$\alpha_c=0$ for adiabatic phonons. 

For polyacetylene Fradkin and Hirsch \cite{fradkin-ssh} used rescaled phonon 
parameters, which in our units correspond to $\omega_0=0.067$, $\alpha=0.052$.
This point is indicated in Fig.~\ref{fig:alphac}, and, in accordance with the 
strong dimerization of polyacetylene, is well within the Peierls phase.

\begin{figure}
\includegraphics[width=8.3cm]{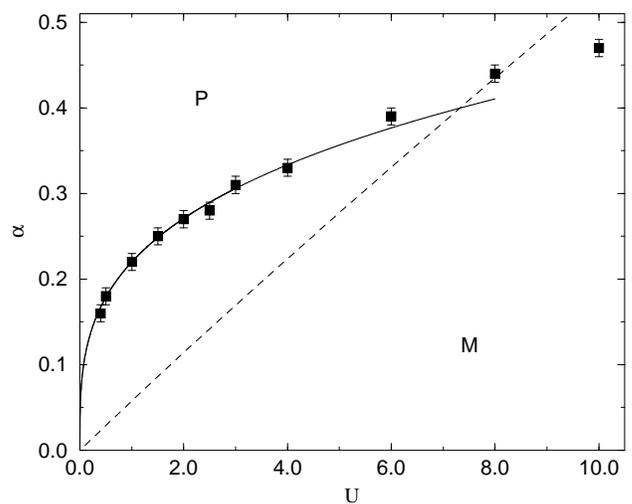}
\caption{Phase diagram for $\omega_0=1, V=U/4$. The squares with
error bars show the critical e-ph coupling. The solid curve shows the form
$\alpha_c \sim U^{0.3}$. The dashed curve shows the transition for the 
corresponding spin-Peierls model.\cite{raas}}
\label{fig:ssh}
\end{figure}

As argued above, the e-ph coupling in the present model is similar to that
in the SSH model for the purpose of studying the dimerization transition.
This implies that in the limit of $(U,V)\rightarrow 0$, we should be able
to reproduce previous SSH results. In particular, according to Fradkin and 
Hirsch,\cite{fradkin-ssh} $\alpha_c$ should be zero even for finite frequency
phonons. To verify this, we have studied $\alpha_c$ as a function of $(U,V)$, 
keeping a fixed ratio of $U/V=4$. With this ratio the ground state for all 
$U$ is a Mott insulator with zero spin gap in the limit of vanishing 
e-ph interaction.\cite{pinaki} We have studied only a single phonon frequency, 
$\omega_0=1$. The resulting phase diagram is presented 
in Fig.~\ref{fig:ssh}. The critical e-ph coupling decreases monotonically 
with decreasing $(U,V)$, but the smallness of the spin gap as $(U,V) \to 0$ 
makes it hard to obtain reliable results below $U=0.4$. We can therefore not 
make a definite statement about this limit. Nevertheless, our results are 
consistent with a power-law behavior $\alpha_c \sim U^\gamma$ with 
$\gamma\approx 0.3$, but a logarithmic form for $U\to 0$ can also not be 
excluded.

For large $U,V$, and with the ratio $V=U/4$, the extended Hubbard model can 
be mapped onto the spin-$\frac{1}{2}$ Heisenberg chain with exchange coupling
$J=4t^2/(U-V)$ (this mapping becomes invalid for $V\approx U/2$, where
phase transitions to bond-ordered and charge-ordered phases occur
\cite{voit,nakamura,pinaki}). Carrying out this transformation for the full 
electron-phonon model (\ref{eqn:H}), the phonon-modulated exchange is
\begin{equation}
J(x_i) = {4t^2 (1 + \alpha x_i)^2 \over U-V},
\end{equation}
where $x_i = a^{\dagger}_i + a_i$. Under the assumption that the non-linear
term $\sim (\alpha x_i^2)$ can be neglected, which is not {\it a priori} clear 
when $\alpha$ is not small,  we obtain exactly the spin-Peierls model 
considered in, e.g., Refs.~\onlinecite{anders-sp} and \onlinecite{raas}:
\begin{equation}
H_{\rm SP} = \sum_{i} (J+g x_i)\mathbf{S}_i \cdot \mathbf{S}_{i+1} +
\omega_0 \sum_i a^{\dagger}_ia_i,
\label{Hsp}
\end{equation}
with
\begin{equation} 
J={4t^2\over U-V},~~~~ g={8\alpha t^2\over U-V} .
\label{Jg}
\end{equation}
For the model (\ref{Hsp}), an analytic expression for the critical spin-phonon
coupling $g$ has been obtained for the whole range of bare phonon 
frequencies $\omega_0/J$.\cite{raas} The form, Eq.~(12) of 
Ref.~\onlinecite{raas}, is expected to be exact in the anti-adiabatic 
limit, $\omega_0/J \to \infty$, which here corresponds to $U\to \infty$. It 
is in good agreement with numerical (SSE) results \cite{anders-sp} for 
the spin-Peierls transition  even for frequencies as low as $\omega_0/J=0.25$.
In Fig.~\ref{fig:ssh} we compare our SSE results for the extended 
Peierls-Hubbard model with the spin-Peierls form for $U$ up to $10$. The
transition curve crosses the Mott-Peierls transition curve at $U\approx 8$,
and is not in good agreement away from this point. The spin-Peierls critical
$\alpha$ increases linearly with $U$ as $U \to \infty$, whereas the 
Mott-Peierls boundary has a slower increase with $U$. The poor agreement 
for small $U$ is not surprising, as the mapping to the spin chain, with 
coupling $J$ given by (\ref{Jg}), can only be expected to be good for large 
$U$, and the form used for the spin-Peierls transition curve is not expected
to be quantitatively accurate for very small $\omega/J$ (corresponding 
here to small $U$).\cite{raas} Considering, however, that the analytical
form {\rm is} accurate for the spin-Peierls model with $\omega_0/J=0.25$,
\cite{raas,anders-sp} and that the nonlinear spin-phonon coupling should
be negligible for small $\alpha$, the disagreement for the small-$U$ region
in Fig.~\ref{fig:ssh} must be due to the poor correspondence between the 
full electron model and the spin chain with the lowest-order $J$ in 
Eq.~(\ref{Jg}). The poor agreement for $U > 8$ indicates that the nonlinear
spin-Phonon term $(\alpha x_i)^2$ {\it does} becomes important as 
$U \to \infty$. 

One effect of the nonlinear coupling term is to renormalize $J$: Writing 
$x_i^2 = \langle x_i^2\rangle + \Delta (x_i^2)$, the renormalized J is given by
\begin{equation} 
J_{\rm eff} = {4t^2(1 + \alpha^2 \langle x_i^2) \rangle\over U-V},
\end{equation} 
i.e., $J_{\rm eff} > J$. Evaluation the spin-Peierls transition curve using 
$J_{\rm eff}$ instead of $J$ clearly would reduce $\alpha_c$ for given $U$ 
and bring the result closer to the actual Mott-Peierls curve in 
Fig.~\ref{fig:ssh}. However, we have not evaluated $\langle x_i^2 \rangle$ 
for a quantitative test of this effect. 
In any case, if $\langle x_i^2 \rangle$ is large there
is also no reason to expect that the remaining nonlinear coupling
$\Delta (x_i^2)$ can be neglected if $\alpha$ is not small. This issue 
clearly deserves further study.

\section{summary}

In summary, we have studied several aspects of the phase diagram of an
extended 1D Hubbard model coupled to optical bond phonons. We have 
demonstrated that the stochastic series expansion technique 
\cite{sseloop,pinaki}
can be used for large electron-phonon chains (here up to 256 sites) to 
compute several different quantities that signal the opening of the spin 
gap at the Mott-Peierls transition. The spin gap boundary is also in good 
agreement with direct probes of the bond order (i.e., kinetic-energy 
correlation functions and susceptibilities at $q=\pi$).

Our phase diagrams are in agreement with what is generally expected, but
to our knowledge they have not been computed quantitatively before. For
large e-e couplings, we have pointed out the relevance of an effective
nonlinear spin-phonon coupling in the mapping of the Hubbard model to a 
spin chain. Because of this, standard spin-phonon models, where the nonlinear
term is not included, cannot be expected to reproduce fully the phase diagrams
of electron-phonon models.

The methods that we have used here should also be applicable to systems
away from half-filling. We plan such calculations aimed at studying the
stability of the soliton lattice, which is formed in doped systems in the 
adiabatic limit,\cite{ssh,salkola,jeckelmann-mi} in the presence of 
finite-frequency phonons. A previous quantum Monte Carlo study has 
indicated that the soliton lattice is stable.\cite{takahashi} However, 
open boundary conditions were used for relatively small chains, and therefore
the issue of whether this is a stable phase on an infinite lattice 
remains to be clarified.

\acknowledgments

This work was supported by the NSF under Grants No.~DMR-99-86948 (PS) and 
DMR-97-12765 (DKC), and by the Academy of Finland, project 26175 (AWS).
The numerical calculations were carried out at the NCSA in Urbana, Illinois.

\end{document}